

\tolerance=1200
\magnification 1200
\baselineskip=12pt plus 1pt
\parindent=25pt

\font\small=cmr10 at 10truept
\baselineskip=20pt plus 1pt


{\nopagenumbers
\hfill UMN-TH-1024/1992

\hfill MAY 1992

\vskip1.5cm
{\bf
\centerline{Stringy Cosmic Strings and Axion Cohomology}
}

\vskip1.5cm

\centerline{{\bf Nemanja Kaloper}$~^{*}$}
{
\small
\baselineskip=12pt plus 1pt
\footnote{}{$^{*}~~$Address after Sep. 1, 1992: Dept. of Physics,
Univ. of Alberta, Edmonton, Alta T6G 2J1, Canada}
}
\centerline{School of Physics and Astronomy}
\centerline{University of Minnesota}
\centerline{Minneapolis MN 55455}
\vskip1.5cm
\centerline{\bf Abstract}

{
The static stationary axially symmetric background ("infinite cosmic string")
of the $D=4$ string theory provided with
an axion charge is studied in the effective theory approach. The most general
exact solution is constructed.
It is found that the Kalb-Ramond axion charge,
present in the string topology $R^{3} \times S^{1}$, produces nontrivial
gravitational field configurations which feature horizons. The corresponding
``no-hair'' theorems are proved which stress uniqueness of black strings.
Connection of the solutions with the gauged WZWN sigma model constructions
on the world sheet is discussed since they are the only target spaces
which hide their singularities behind horizons, and thus obey the cosmic
censorship conjecture.
}

\vfil
\eject}

{\bf Introduction}
\vskip1.5cm

Interest in cosmic strings as possible gravitational solutions has arisen
in response to the study of cosmological phase transitions in the early
Universe. Investigations of mechanisms of the phase transitions have led
us to believe that the field configurations specific to theories with
spontaneously broken symmetries can provide the necessary energy-momentum
sources to support string-like solutions of the coupled gravity-matter
equations$^{1)}$. Two generic types of solutions were found, depending
on the type
of the symmetry which gets broken. When the spontaneously broken symmetry
is a local gauge symmetry, the string solution associated with it appears
to be devoid of all long-range interactions, due to the
low energy cutoffs introduced by the symmetry breaking$^{1,2)}$
(except, possibly,
in case of the superconducting cosmic string, which may have electromagnetic
long-range interactions with interstellar plasma$^{3)}$). Thus, the only
non-trivial gravitational effect produced by a local string is the deficit
angle, a phenomenon truly topological in its nature.

The situation is
significantly different in case of global strings. These were shown to be
consistent gravity-matter configurations in theories with broken global
symmetries. Long-range gravitational interactions persist in this case due
to the coupling of gravity to Goldstone bosons left after the symmetry
breaking. Presence of the Goldstone bosons actually assures there are
no low energy cutoffs that rule out long-range interactions for local strings.
However, for the very same reason the gravitational field associated with
global strings is not asymptotically flat, since the energy-momentum density
of the matter does not fall off to zero rapidly enough. The gravitational
force of the string, curiously, turns out to be repulsive$^{4-6)}$. A
peculiarity of presence of a horizon in the geometry of a global string was
observed by Harari and Polychronakos$^{7)}$ in the limit of small Goldstone
boson energy density (necessary in their analysis for establishing boundary
conditions at the string core), and further elucidated by Gibbons et al$^{8)}$.
In particular, the analysis of Ref. [ 8 ] points out general circumstances
under which global cosmic strings exist, which are singular and possess event
horizons. It is interesting to note that in order to find such solutions,
Gibbons et al. resort to string theoretic sigma model target spaces as matter
sources.

Stringy aspects of cosmic strings have already been investigated$^{9-11)}$,
along with cosmic strings in other nonminimal theories of
gravity$^{12)}$. A common feature of such treatises is that all these
theories are characterized with the presence of
an extra scalar field non-minimally
coupled to gravity and/or matter, so that its variation produces non-trivial
gravitational field exterior to a string even for local cosmic strings.
Of particlular interest here are the recent studies which attempt to extend
the $1+1$ Witten black hole solution$^{13)}$ and its conformal field theory
interpretation to higher dimensions. Some solutions of the gauged
Wess-Zumino-Witten-Novikov (WZWN) sigma
models on axially symmetric static target
spaces have been constructed already$^{14-18)}$. More comprehensive
approaches are under way$^{19)}$.
Also, investigations of topological configurations in the massless sector
of the theory have been conducted, leading to supersymmetric instanton
solutions$^{20)}$.

In this paper a class of exact solutions with an
axion charge will be derived and
analyzed. It will be demonstrated that classically they represent global
string-like solutions with different analytic properties due
to the presence of a nontrivial dilaton field
and axion charge around it. A black string
version of the ``no-hair'' theorem will be proved, showing uniqueness
of a class of charged $~4D~$ black strings with horizons.
Those are singled out, as they correspond to the WZWN models
and represent exact conformal field theories on the string worldsheet.
As such, they are viable candidates for the string field theory vacua.

\vfil

\eject
{\bf Axion Cohomology}
\vskip1.5cm

The starting point of our investigation is the action for the bosonic sector
of the supergravity multiplet in the background field formulation of string
theory. In the Einstein frame (in all that follows Yang-Mills fields are
ignored) to order $O(\alpha^{0})$ it is

$$
S=\int d^{4}x\sqrt{g}\big({1 \over 2\kappa^{2}}R - e^{-2\sqrt{2}\kappa \Phi}
H_{\mu\nu\lambda}H^{\mu\nu\lambda}-{1 \over 2} \partial_{\mu}\Phi
\partial^{\mu}
\Phi \big) \eqno(1)
$$

\noindent where $R$ is the Ricci scalar, $H_{\mu\nu\lambda}$ the Kalb-Ramond
axion field strength and $\Phi$ is the dilaton. Conventions of Ref. [ 22 ]
are followed throughout the paper.
Here it was implicitly assumed that the cosmological constant at the string
world-sheet level, leading to the exponential dilaton potential in the
Einstein frame action, is zero. This assumption can only be justified by
inspecting evolution of the structure of vacuum with the complete
description of string theory. In the absence of such a description, it is
reasonable to imagine that the various conformal anomalies conspire to
yield cancellation of the total central charge (for example, the
string vacuum could be decomposed as
$M^{4}\times K^{6}$ where the internal degrees of freedom yield the
cancellation of the central charge, but at low enough energies are
dynamically decoupled). Furthermore, the compactification scale is
expected to be high above the supersymmetry breaking scale where the dilaton
obtains mass. Hence solutions of the theory above may be expected to describe
cosmic strings in an early phase of the Universe, somewhere between
the compactification and the supersymmetry breaking scales.
For the sake of simplicity, we will first ignore the dilaton field and
look only at the coupled axion-gravity system. It is our goal here to examine
the influence of the nonvanishing axion charge on the gravitational field.

The metric of a cosmic string is described by
the cyllindrically symmetric static stationary background,

$$
ds^{2}=-e^{2\nu}dt^{2}+e^{2\lambda}dr^{2}+e^{2\mu}dz^{2}+e^{2\eta}d\phi^{2}
\eqno(2)
$$

\noindent where the metric functions depend only on the radial coordinate $r$.
The four metric functions are not independent, however, since by a coordinate
transformation (for example, $\rho=e^{\eta (r)}$) any combination of them can
be fixed. It is convenient to keep all four arbitrary for calculational
purposes and simplify the equations of motion by a specific choice. This
freedom is a remnant of the full $GL(3,1)$ gauge invariance of gravity in
$4D$.

The equations of motion are, in terms of the dual of the Kalb-Ramond
field strength,

$$
V_{\mu}=\sqrt{g}\epsilon_{\nu\lambda\sigma\mu}H^{\nu\lambda\sigma}
\eqno(3)
$$

\noindent of the form (in $\kappa^{2}=1$ units)

$$
R_{\mu\nu}-{1 \over 2}g_{\mu\nu}R={1 \over 3} V_{\mu}V_{\nu}-
{1 \over 6}g_{\mu\nu}V^{2}
\eqno(4)
$$

\noindent and

$$
dV=d~{^{\ast}V}=0  \eqno(5)
$$

\noindent where $V=V_{\mu}dx^{\mu}$ is the associated 1-form.

Usually, one would search for

$$
V=da \eqno(6)
$$

\noindent as the solution of the two axion equations. In the most
general case when the base manifold has nontrivial first cohomology,

$$
V=da+C_{A}\gamma^{A} \eqno(7)
$$

\noindent where $\gamma^{A}$ are the harmonic forms
generating the first cohomology
group of the manifold and $C_{A}$ are constants. Upon substitution
of the axion solution in the Einstein's equations for the assumed background
describing a base manifold of topology $R^{3}\times S^{1}$ one sees that the
axion can be represented by a purely topological contribution$^{3)}$

$$
V=Q~d\phi \eqno(8)
$$

\noindent where $Q$ is the axionic charge per unit length of the string:

$$
Q={1 \over 2\pi}\int_{S^{1}}V \eqno(9)
$$

\noindent This charge generalizes the uniform line distribution of electric
charge, as is obvious from $Q \sim \int_{S^{1}} {^{\ast}H}$.

Clearly, due to its topological nature expressed via the Gauss law
above, the charge $Q$ is conserved. Furthermore,
we can rewrite the energy-momentum tensor of this configuration as

$$
T_{\mu\nu}=~{Q^{2} \over 6}~e^{-2\eta}~diag (1, -1, -1, 1) \eqno(10)
$$

\noindent recognizing it as the matter source giving rise to the gravitational
field of a global cosmic string. This should not be a surprise, since
Gibbons et al.$^{8)}$ showed that global strings arise whenever matter sources
can be identified with target spaces containing closed geodesics. As a
consequence, gravitational field of global strings arises when the principal
bundle describing matter content is of nontrivial first cohomology. In the
case demonstrated above where $~V=Qd\phi~$ one can identify  the space-time
angle $\phi$ as a representation of a matter field living on a $U(1)$
target manifold, a closed target space geodesic by itself. Then$^{1,2,4,5)}$
it is easy to write down the solution for the gravitational field around such a
cosmic string:

$$\eqalign{
ds^2=~&-({r \over r_{0}})^{1-w}~dt^{2}~+~({r \over r_{0}})^{1+w}~dz^{2}\cr
{}~&+~{\Delta}^{2}({r \over r_{0}})^{{w^{2}-1 \over 2}}
\exp{(-{Q^{2} \over 6}({r \over r_{0}})^{2})}dr^{2}
{}~+~{\Psi}^{2}({r \over r_{0}})^{{w^{2}-1 \over
2}} \exp{(-{Q^{2} \over 6}({r \over r_{0}})^{2})}d\phi ^{2}\cr}  \eqno(11)
$$

\noindent Here $w,~r_{0},~\Delta$ and $\Psi$ are constants of integration.
Observe, that when $Q~=~0$ the solution above correctly reduces to the
static cyllindrically symmetric metric solving the vacuum Einstein's equations,
i.e., to the class of solutions containing the local cosmic
string (to demonstrate
this one would have to take the zero limit of other parameters
in $(11)$ as well).
With the help of a coordinate transformation, the case $~w=-1~$
can be cast
in the form first discussed by Harari and Polychronakos$^{4)}$,

$$\eqalign{
ds^2=~&-(1-{1 \over \sigma} \ln{({\rho \over \rho_{0}})})~d\tau^{2}~+~dz^{2}\cr
{}~&+(1-{1 \over \sigma} \ln{({\rho \over\rho_{0}})})^{-1}~d\rho^{2}
{}~+~{Q^{2} \sigma \over 3} \rho^{2}~d\phi ^{2} \cr} \eqno(12)
$$

\noindent where $\sigma~=~\ln{\Delta}$ and
$\rho_{0}{}^{2}~=~{3r_{0}{}^{2} \over Q^{2} \sigma}$. The Kalb-Ramond axion
then is

$$
H=~{1 \over 2 Q \sigma \rho^{2}}~d\tau \wedge dz \wedge d\rho
\eqno(13)
$$

\noindent Note that in the limit $~Q \rightarrow 0~$,
$~\sigma \rightarrow \infty~$, $~Q^{2} \sigma = const.~$ equations (12-13)
reduce to the local cosmic string, as claimed above.

The difference
between this solution and the one  previously analyzed  by Harari and
Polychronakos is that it is an exact, everywhere valid solution
representing a doubly singular line distribution of axial charge, the
singularities being located at the origin and at infinity. Note, however, the
strange property of the above solution, that if we allow imaginary
axion charge, and at the same time flip the sign of $\sigma$,
we obtain the (more appealing) configuration where
the physical region of the metric
$(12)$ is outside of the horizon $~\rho_{h}=\rho_{0} \exp{\sigma}~$,
as opposed to the solution of Harari and Polychronakos, who find a rather
peculiar situation where the physical region of the metric is inside
it. In this case, the metric $(12)$ has a well defined Newtonian limit,
the gravitational potential being

$$
\Phi_{N}=~-{1 \over 2\mid \sigma \mid}~\ln{\rho \over \rho_{0}} \eqno(14)
$$

\noindent exactly the expression for the
Newtonian potential of a nonrelativistic
line distribution of mass with density per unit length
$~2 \pi {\mid \sigma \mid ^{-1}}$.
Then it is easy to understand the appearance of the singularity at
infinity. This is just the usual infrared divergence in the classical limit of
a theory describing massless particles.

It is interesting to note that the need for an imaginary axion to produce
the solution $(12)$ with $\sigma >0$ can be relaxed
if the three form $H_{\mu\nu\lambda}$ has
been given a tachyonic kinetic term instead. Such a degree of freedom might
arise, for example, after dimensional reduction of a higher rank form defined
in some background metric with compactified internal space of pseudoeuclidean
signature.
\vskip1.5cm
{\bf The Supergravity Multiplet}
\vskip1.5cm
We will now investigate the cylindrically symmetric static solutions with
axionic charge on the target manifold of topology $R^{3} \times S^{1}$ in
the theory describing the complete bosonic sector of the supergravity
multiplet.
It contains the graviton, the dilaton and the three form Kalb-Ramond axion.
This time, however, we choose to work in the world sheet frame, which  has been
argued to represent the more natural background for discussing the  string
propagation in curved backgrounds, its role being exactly the string coupling
constants. Also, this will be more convenient for later comparison  with the
WZWN constructions. Of course, the analysis of the
background field equations of motion, understood as a
classical theory, really does not  depend on which frame
one  chooses to work in, since a simple conformal transformation relates the
equations of motion in different frames.

Thus, the action describing the theory on the target space is, to order
$O(\alpha^{0})$ in $4D$,

$$
S=\int d^{4}x\sqrt{G}  e^{-\sqrt{2}\kappa \Phi} \big({1 \over 2\kappa^{2}}R -
H_{\mu\nu\lambda}H^{\mu\nu\lambda}+
\partial_{\mu}\Phi \partial^{\mu} \Phi + \Lambda \big) \eqno(15)
$$

\noindent where the cosmological constant has been included for the background
theory above to represent the noncritical string theory too. The world
sheet cosmological constant can arise from the leftover conformal anomaly
which may not have been cancelled between the central charges of the particle
and ghost systems. Then, the requirement of cylindrical symmetry of the
target space is equivalent to using the metric (2) on the world sheet and
allowing  the dilaton to depend on the radial coordinate only. As for the
axion,
since we are interested in picking it so that it is given by the $S^{1}$
cohomology  of the manifold, its equations of motion $~dH=0~$ and
$~d\exp{(-\sqrt{2}\kappa \Phi)}{^{*}H}=0~$ in conjunction with the duality
transformation and the metric ans\" atz, yield

$$\eqalign{
B~=&2B_{20}~dz \wedge dt \cr
H~=&2B_{20}'~dr \wedge dz \wedge dt  \cr
B_{20}'=&-{Q \over 2}~
\exp{(\sqrt{2} \kappa \Phi + \nu + \mu + \lambda - \eta)}\cr}
\eqno(16)
$$

\noindent where the prime denotes a derivative with respect to $r$. The
constant of integration $Q$ represents the axion cohomology charge, defined
via the Gauss law $(9)$, which in this case can be writen as

$$
Q~=~{1 \over 2 \pi} \int_{S^{1}}~e^{-\sqrt{2} \kappa \Phi}~{^{*}H} \eqno(17)
$$

It is easiest to work in the action on the tangent bundle to derive the
equations of motion. Defining the locally flat coordinate system with tetrads
$~e^{\alpha}= \exp{(\nu_{\alpha})}dx^{\alpha}~$ (no summation) with
$~\alpha \in \{ 0, \ldots ,3 \}~$ and computing the connexion and curvature
forms, the action can be rewritten in terms of the degrees of freedom, in units
of Planck's mass $~(\kappa^{2}=1)~$, as

$$\eqalign{
S~=~\int~dr \Bigl\{~& \Bigl( \nu' \mu' + \nu' \eta' + \mu' \eta' -
\sqrt{2} \Phi' \bigl( \nu' + \mu' + \eta' \bigr) \Bigr) e^{ \nu + \mu + \eta
- \lambda - \sqrt{2} \Phi}\cr
&+ {{\mit \Upsilon}^{2} \over 6}~e^{ \eta - \nu - \mu - \lambda -
\sqrt{2} \Phi}\cr
&+ \Lambda~ e^{ \nu + \mu + \eta + \lambda - \sqrt{2} \Phi}
+ \Phi'^{2}~ e^{ \nu + \mu + \eta - \lambda - \sqrt{2} \Phi} \Bigr\} \cr}
\eqno(18)
$$

\noindent where $~{\mit \Upsilon}~=~Q \exp{( \nu + \mu + \lambda - \eta +
\sqrt{2} \Phi)}~$ is obtained after substitution of the axion solution (16).

Variation of the action $(18)$ gives the equations of motion. With some simple
algebra, and after the choice of gauge $~\lambda~=~\nu + \mu - \sqrt{2} \Phi~$
has been made, they become:

$$\eqalign{
&\eta'' +~\eta'^{2}~=0 \cr
\mu''~-~&\nu'' +~(\mu' - \nu')~\eta'~=~0 \cr
\mu''~+~\nu'' +~&(\mu' + \nu')~\eta'~=~{2 \over 3}
Q^{2}~e^{2(\nu + \mu - \eta)} \cr
\Phi''~+~\Phi'~\eta'~=
{}~{\sqrt{2} \over 3}&Q^{2}e^{2(\nu + \mu - \eta)}~-~\sqrt{2}\Lambda
{}~e^{2(\nu + \mu - \sqrt{2} \Phi)} \cr
\nu'\mu' + \nu'\eta' + \mu'\eta' - \sqrt{2} \Phi' (\nu' + &\mu' + \eta')
+ \Phi'^{2} = \Lambda e^{(\nu + \mu - \sqrt{2}\Phi)}
+ {Q^{2} \over 6} e^{(\nu + \mu - \eta)}\cr}
\eqno(19)
$$

The system of equations above is exactly integrable. Clearly, the function
$\eta$ serves the role of the "evolution kernel" and
is determined by a Riccati equation
which in this case is easy to solve. There exist two different classes of
solutions, determined as follows. The general (inequivalent) solutions of the
Riccati equation are $~\eta~=~\eta_{0}~$ or
$~\eta~=~\eta_{0}~+~\ln{(r + r_{0})}~$
where $\eta_{0}$ and $r_{0}$ are integration constants. The constant $r_{0}$
is irrelevant and may be dropped. The other constant, however, may be
physically significant as it measures the deficit angle.
The remaining equations
can be simplified upon substitution of these formulas and integrated.
Hence, the solutions are

$${~}$$
\eject

\noindent Case 1. :

$$\eqalign{
&~~~~\eta~=~\eta_{0}  \cr
{}~~~~\mu~&-~\nu~=~Y_{0} + Y_{1}~r \cr
&\mu~+~\nu~=~X \cr
&\sqrt{2}\Phi~=~X - W \cr
&\cr
X'~=~\pm~& \sqrt{2C - Y_{1}{}^{2} + {2 \over 3}Q^{2} \exp{2(X - \eta_{0})}} \cr
W'~=~&\pm~ \sqrt{C + 2\Lambda \exp{(2W)}} \cr}
\eqno(20)
$$

\noindent and

\noindent Case 2. :

$$\eqalign{
&~~~~~~\eta~=~\eta_{0} + \ln{ r} \cr
&~~\mu~-~\nu~=~Y_{0} + Y_{1}~\ln{ r} \cr
&~~~~~~\mu~+~\nu~=~X \cr
&~~\sqrt{2} \Phi~=~X + \eta - W \cr
&\cr
X'~=~\pm~& { \sqrt{2(C - 1) - Y_{1}{}^{2} +
{2 \over 3} Q^{2} \exp{2(X - \eta_{0})}} \over r } \cr
W'~=~&\pm~{ \sqrt{C + 2\Lambda \exp{2(W - \eta_{0})}} \over r } \cr}
\eqno(21)
$$

\noindent Here $~\eta_{0},~Y_{0},~Y_{1}~$ and $~C~$, together with
$~\nu_{0}~$ and
$~\mu_{0}~$ are the (independent) integration constants which
completely determine
solutions. We will always assume $~Q\ne 0~$, as we are interested in the
effects of the axion cohomology. Besides, the case $~Q=0~$ is equivalent
to the Jordan-Brans-Dicke cosmic string of ref. [ 12 ]. Furthermore, we
will require that
$~\Lambda=-{2 \over 3}\delta c_{T} = {2 \over 3}(c_{T}-4) \ge 0~$,
as is known to be the case in string theory$^{22)}$. The central
charge deficit $~\delta c_{T}~$
arises via the algebraic construction of the target
manifold. The $~\lambda~$ is determined then with the gauge choice above.

Typically,
the integrals above are combinations of polynomials and
hyperbolic and trigonometric
functions. The trigonometric functions, however, are highly undesirable,
since their
periodicity compactifies the radial direction and introduces an infinite
number of
ring-like singularities inconsistent with the assumed
cylindrical structure of the
target space. Avoiding them is guaranteed with choice $~C=g_{0}{}^{2}$, and
$~2 g_{0}{}^{2} - Y_{1}{}^{2}=f_{0}{}^{2}~$ or
$~2 g_{0}{}^{2} - Y_{1}{}^{2} - 2=f_{0}{}^{2}~$, respectively.

The explicit solutions can be classified with respect to the values of
$~C,~Y_{1},~Q~$ and $~\Lambda~$. They are listed below.

\noindent Strings with $~\Lambda=0~$:

$$\eqalign{
ds^{2}~=~&-\sqrt{3 \over 2} {f_{0} e^{ \eta_{0}} \over Q}
{\bigl( e^{-Y_{1}r} dt^{2} - e^{Y_{1}r} dz^{2} \bigr) \over \sinh(A \mp
f_{0}r)}
+ e^{2W_{0} \pm 2g_{0}r} dr^{2} +
e^{2 \eta_{0} + 2 \sigma r}d\phi^{2} \cr
&e^{-\sqrt{2} \Phi}~=~\sqrt{2 \over 3} {Q e^{-\eta_{0}} \over f_{0}}
e^{W_{0} \pm g_{0}r - \sigma ( \eta_{0} + r)} \sinh(A \mp f_{0}r) \cr
&~~~~~B_{20}~=~D \pm {3 \over 4} {f_{0} \over Q} e^{(1 + \sigma) \eta_{0}}
\coth(A \mp f_{0}r) \cr}
\eqno(22)
$$

\noindent Strings with $~\Lambda >0~$:

$$\eqalign{
ds^{2}~=~-\sqrt{3 \over 2} &{f_{0} e^{ \eta_{0}} \over Q}
{\bigl( e^{-Y_{1}r} dt^{2} - e^{Y_{1}r} dz^{2} \bigr) \over \sinh(A\mp f_{0}r)}
+ {g_{0}{}^{2} \over 2 \Lambda} {dr^{2} \over \sinh^{2}(B \mp g_{0}r)}
 + e^{2 \eta_{0} + 2 \sigma r}d\phi^{2} \cr
&e^{-\sqrt{2} \Phi}~=~{Q e^{-\eta_{0}} g_{0} \over f_{0} \sqrt{3 \Lambda} }
{\sinh(A \mp f_{0}r) \over \sinh(B \mp g_{0}r)}
e^{-\sigma ( \eta_{0} +  r)} \cr
&~~B_{20}~=~D \pm {3 \over 4} {f_{0} \over Q} e^{(1 + \sigma) \eta_{0}}
\coth(A \mp f_{0}r) \cr}
\eqno(23)
$$

\noindent with
$~2g_{0}{}^{2}~=~f_{0}{}^{2} + Y_{1}^{2} +2\sigma~$. The parameter
$~\sigma~$ can take values 0 and 1.
A coordinate transformation has been performed to bring the solution of the
second case to the form above. Note that in case of $~\Lambda~$ of opposite
sign the only change would be replacing $~\sinh(B \mp g_{0}r)~$ with
$~\cosh(B \mp g_{0}r)~$.

The coordinate frame chosen for representing the solutions above is
very transparent for further analysis.
We note a few interesting features of these configurations.
It is not difficult to evaluate the curvature invariants of (22-23) and
inspect the type of potential singular behaviour in the metric (we absorb
$~\pm~$ in the parameters).

$$\eqalign{
R~=~-2e^{-2W_{0} + 2g_{0}r} \Bigl(&{7 \over 4}
{f_{0}{}^{2} \over \sinh^{2}(A + f_{0}r)}
- f_{0}g_{0} \coth(A + f_{0}r) + {f_{0}{}^{2} + g_{0}{}^{2} \over 2} \cr
&+ \sigma (f_{0} \coth(A + f_{0}r) + g_{0}) + \sigma^{2}
- {\sigma \over 2} \Bigr) \cr}
\eqno(24)
$$

\noindent and

$$\eqalign{
R~=~- {4 \Lambda \over g_{0}{}^{2}}& \sinh^{2}(B + g_{0}r)
\Bigl({7 \over 4}
{f_{0}{}^{2} \over \sinh^{2}(A + f_{0}r)}
- f_{0}g_{0} \coth(A + f_{0}r) \coth(B + g_{0}r) \cr
&+ {f_{0}{}^{2} + g_{0}{}^{2} \over 2}
+ \sigma ( g_{0}\coth(B + g_{0}r) - f_{0} \coth(A + f_{0}r))
+ \sigma^{2} - {\sigma \over 2} \Bigr) \cr}
\eqno(25)
$$

We will first concentrate on the noncritical string case.
Assume $~g_{0} A \ne f_{0} B~$. Then, the solutions
contain a singularity determined by $~r_{s}= -{A \over
f_{0}}~$. Hence they are charts of the target valid only on open intervals
$~r \in (-\infty, r_{s})~$ and $~r \in (r_{s},\infty)~$.
On the other hand, the surface $~r_{\infty}= -{B \over g_{0}}~$
is well behaved, although the coefficient of $~dr^{2}~$ diverges here.
The curvature actually vanishes here. Inspection of (25)
leads to the conclusion that in order to prevent
blowing up of the curvature at ``spatial infinities'', one has to require
$~f_{0}-g_{0}=\pm \sigma~$ (with the ``regular'' point at $~\pm \infty~$).
Note that this is just a necessary but not sufficient condition for
regularity. That is the reason for the quotation marks. Regularity at the point
would still have to be doublechecked. Only when $~\sigma=0~$ will both
``spatial infinities'' be nonsingular. Regularity of one implies regularity
of the other. When $~\sigma=1~$, there appears another
singularity at one of the ``infinities''.

The constraint $~2g_{0}{}^{2}~=~f_{0}{}^{2} + Y_{1}^{2} +2\sigma~$
translates into  $~f_{0}{}^{2} \mp 4 \sigma f_{0} = Y_{1}^{2}~$. When
$~\sigma=0~$ it is indeed very stringent: it asserts that
$~f_{0}=g_{0}=\pm Y_{1}~$. This fixes the solution uniquely. It is constructed
as follows; for simplicity, assume that all the parameters are positive, and
that $~f_{0} B < g_{0} A~$.
This does not restrict generality of the construction.
Take the patch $~(r_{\infty}, \infty)~$ to describe the exterior of the
black string; identify $~r_{\infty}~$ with the
physical spatial infinity, and $~\infty~$ with the (outer) horizon.
Note that the choice of $~r_{\infty}~$ as the physical spatial infinity is
motivated by the vanishing curvature there. It tells that such space-time is
asymptotically flat. The identification of $~\infty~$ as the (outer) horizon
is consistent as the curvature is finite there, although the metric appears
singular. This signals the horizon. Also note that the singularity at $~r_{s}~$
is pushed ``outside'' of the space-time, and that asymptotic flatness
guarantees that it is completely decoupled from the physical region of the
space-time. Similarly, construct the interior from the patch
$~(-\infty, r_{s})~$: flip the roles of the $~t~$ and $~z~$ coordinates,
identify $~-\infty~$ with the inner horizon, and note
that then $~r_{s}~$ represents a singularity in the space-time, which is
unavoidable because the point $~r_{\infty}~$ was imbedded in the outside.
Finally,  interpolate the region between the horizons with a Bianchi I type
universe, equivalent to the region of imaginary $~r~$
and the parameters $~g_{0}, f_{0}, Y_{1}~$ in the original
coordinates. This patchwork can be represented compactly with a transformation
of  coordinates, and the analytic continuation spelled out above. The
transformation is $~\rho=g_{0} \coth(B + g_{0}r)~$,
with $~f_{0}=g_{0}=Y_{1}~$, and the solution becomes, with the help
of additional simple coordinate transformations,

$$\eqalign{
ds^2~=~-(1-{x_{+} \over x})~dt^2 + &(1-{x_{-} \over x})~dz^2
+ {dx^2 \over 2\Lambda(x - x_{+})(x - x_{-})}  + e^{2 \eta_{0}}d\phi^2 \cr
&e^{-\sqrt{2}\Phi}~=~{Qe^{-\eta_{0}} \over \sqrt{3 \Lambda}}
{x \over \sqrt{x_{+} x_{-}}} \cr
&B_{20}~=~D' - {\sqrt{6} \over 4} { \sqrt{x_{+} x_{-}} \over x} \cr}
\eqno(26)
$$

\noindent where $~x_{\pm}~$ denote locations of the horizons, and
$~x \in (0, \infty)~$.

This is exactly the gauged WZWN sigma model solution first constructed in 4D
by Raiten$^{16)}$ (with a slight modification, the
one-point compactification of the $~\phi~$ space). The limit
when $~f_{0} \rightarrow 0~$ corresponds
to the critical case of the above solution.

In all other cases, when at least one of the ``spatial infinities'' is
a singularity, the situation is dramatically different. For example, suppose
again that the parameters are positive, that $~f_{0} B < g_{0} A~$
and that $~-\infty~$
is a singularity. The patch $~(-\infty, r_{s})~$ then can not be extended
continuously by gluing extra pieces to it since both end points are
singularities. Assuming that the manifold is connected establishes this as
an entire space-time which contains two naked singularities, one at the
origin and the other at infinity. The other patch, $~(r_{s}, \infty)~$,
contains the point $~r_{\infty}~$. If $~\infty~$ were not a singularity
$~(\sigma = 1)~$, it could in principle be made
singularity-free with the point $~r_{\infty}~$
identified with the physical spatial
infinity, and $~\infty~$ with a horizon (or the coordinate origin).
However, closer scrutiny reveals that unless $~f_{0} \in \{ 0,1 \}~$,
$~\infty~$ is a singularity, since the Riemann
curvature squared diverges otherwise
($~\lim_{r\to \infty}{R_{\mu \nu \lambda \sigma} R^{\mu \nu \lambda \sigma}}
\propto f_{0}(f_{0} - 1) \sinh^{4}(B + f_{0}r)~$).
The solution contains
a naked singularity at the origin, and can not be extended past it.
For the remaining two values for $~f_{0}~$, the constraints
$~2g_{0}{}^{2}=f_{0}{}^{2}+Y_{1}{}^{2}+2~$ and $~g_{0}+1=f_{0}~$ rule out
$~f_{0}=1~$. Thus, $~f_{0}=Y_{1}=0~$ and $~g_{0}{}^{2}=1~$; the solution in
this case is

$$\eqalign{
ds^{2}~=~-\sqrt{3 \over 2} &{ e^{ \eta_{0}} \over Q}
{dt^{2} - dz^{2} \over A + r}
+ {dr^{2} \over 2 \Lambda \sinh^{2}(B + r)}
 + e^{2 \eta_{0} + 2 r}d\phi^{2} \cr
&e^{-\sqrt{2} \Phi}~=~{Q e^{-\eta_{0}} \over \sqrt{3 \Lambda} }
{(A + r) \over \sinh(B + r)}
e^{-( \eta_{0} +  r)} \cr
&~~B_{20}~=~D - {3 \over 4} { e^{2 \eta_{0}} \over Q}
{1 \over A + r} \cr}
\eqno(27)
$$

The distinct points of this metric are $~r=-A, \pm \infty, -B~$.
Investigation of the Ricci curvature (25) for this metric then shows that
$~r_{s}=-A~$ and both $~\pm \infty~$ are singularities. So, any one of
the patches $~(-\infty, r_{s})~$, $~(r_{s}, r_{\infty})~$ or
$~(r_{\infty}, \infty)~$ contains at least one singularity. The fact that
the signature of the metric does not change as $~r~$ passes through
$~r_{\infty}~$ says again that $~r_{\infty}~$ can at best be
identified with the physical spatial infinity but is not a horizon.
More properly, a timelike Killing vector transported along a
geodesic passing through  $~r_{\infty}~$ does not flip into
spacelike. Thus, the solution (27) always describes space-times with naked
singularities.

The solutions where either both ``spatial infinities'' are singular, or with
the point $~r_{\infty}~$ sandwiched between $~r_{s}~$ and the singular
``spatial infinity'' are all hopeless, since by necessity they contain naked
singularities.

When $~g_{0} A = f_{0} B~$($~=0~$,by a shift of $~r~$),
the singularity $~r_{s}~$ disappears. Instead,
$~r_{\infty ^{\pm}}=0^{\pm}~$ are regular points. If $~\pm \infty~$ are also
required to be regular, $~f_{0}-g_{0}=\pm \sigma~$. When $~\sigma=0~$, the
solution is fixed uniquely, with both ``spatial infinities'' regular. Hence
both patches $~(-\infty , 0^{-})~$ and $~( 0^{+},\infty)~$ represent
singularity free regions. In effect, they are exact replicas of each other as
can be clearly seen under the transformation
$~r \rightarrow -r,~t \leftrightarrow z~$. Note that the patches have constant
Ricci curvature $~R=-3\Lambda~$ and dilaton
$~\exp{(-\sqrt{2} \Phi)}={\mid Q \mid e^{-\eta_{0}} / \sqrt{3\Lambda}}~$.
Actually, one can compute that, at the tangent bundle, the Riemann curvature
tensor for this configuration is constant, and so are all the curvature
invariants. This solution is completely singularity-free. It is a 3D
anti-de-Sitter space-time crossed with a flat circle. Each patch separately
is an extremal black string of (26), as shown in the 3D case by Horne and
Horowitz$^{15)}$.
Then, to
patch up the solutions, one can identify
$~0^{\pm}~$ with the physical
spatial infinities, and $~\pm \infty~$ with the outer and inner horizons and
interpolate between them with a Bianchi I universe. This in fact again
corresponds  to an analytically continued coordinate transformation. So,
start with  $~\rho =g_{0} \coth(g_{0}r)~$ and rewrite the solution as

$$\eqalign{
ds^{2}~=~-\sqrt{3 \over 2} { e^{ \eta_{0}} \over Q}
\Bigl((\rho - g_{0})~dt^{2} - &(\rho + g_{0})~dz^{2}\Bigr)
+ {d\rho^{2} \over 2 \Lambda (\rho - g_{0})(\rho + g_{0})}
 + e^{2 \eta_{0} + 2 r}d\phi^{2} \cr
&e^{-\sqrt{2} \Phi}~=~{\mid Q \mid e^{-\eta_{0}} \over \sqrt{3 \Lambda} } \cr
&B_{20}~=~D - {3 \over 4} { e^{2 \eta_{0}} \over Q} \rho \cr}
\eqno(28)
$$

Notice that the solution suggests that the base space consists of
two identical images $~\rho \rightarrow -\rho, t \leftrightarrow z~$
glued together along the boundary $~\partial_{\rho=0}M~$.
This is a rather awkward situation, in that one can imagine an observer
moving along a radial geodesic who can enter the replica universe
passing through the axis of the string. More can be learned from
the study of radial geodesics.
The geodesic equations can be
easily integrated to yield (discarding the flat direction $~\phi~$)

$$\eqalign{
&t'~=~K e^{-\rho} \cr
&z'~=~L e^{-\rho} \cr
\rho'^{2}~=~N(\rho - g_{0})(\rho + g_{0}) +
&\sqrt{3 \over 2}{ e^{ \eta_{0}} \over \mid Q \mid}
\Lambda (K^{2}-L^{2})(\rho - g_{0})(\rho + g_{0})e^{-2\rho} \cr}
\eqno(29)
$$

This states that no matter where the observer starts from, he/she must
stop moving in the radial direction when $~\rho \rightarrow g_{0}~$.
In other words, it seems that the point $~g_{0}~$ represents
a classical turning
point where kinetic energy vanishes and motion ceases.
Yet, a simple transformation of coordinates $~x^{2}=\rho - g_{0}~$ confirms the
naive expectation that the passage is possible. Then,

$$
4x'^{2}~=~N(x^{2} + 2 g_{0}) + \sqrt{3 \over 2}
{ e^{ \eta_{0}} \over \mid Q \mid}
\Lambda (K^{2}-L^{2})(x^{2} + 2 g_{0})e^{-2\rho}
\eqno(30)
$$

\noindent and therefore the observer, very slowly, and never passing $~g_{0}~$
in the original coordinates, crosses the horizon. It appears that he/she
can travel all the way past, squeezing through the origin into the region
$~\rho < 0~$ and eventually reenter the copy universe.
Similar tour through the black hole interior was investigated by Horne and
Horowitz in 3D$^{15)}$, with the difference that their observer had only one
horizon to cross. That would correspond to our case when $~g_{0}=0~$.

If $~\sigma=1~$, then both ``spatial infinities'' turn out to be singular,
unless $~g_{0}{}^{2}=1~$ (by the already mentioned asymptotic behaviour of the
square of Riemann tensor). If $~g_{0}{}^{2}=1~$ ($~f_{0}=Y_{1}=0~$),
the metric is the same as (27) with $~A=B=0~$:

$$
ds^{2}~=~-\sqrt{3 \over 2}{e^{-\eta_{0}} \over Q} { dt^2 - dz^2 \over r}
+ {dr^2 \over 2\Lambda \sinh^{2}r} + e^{2(\eta_{0} + r)}~d\phi^{2}
\eqno(31)
$$

\noindent where both $~\pm \infty~$  are again singular.
Consequently, the patches $~(-\infty, 0)~$  and
$~(0,\infty)~$ both contain naked singularities. However, the situation here
is more resemblant of the Harari and Polychronakos solution$^{7)}$,
with an additional bonus that it now possesses
manifest boost invariance in the $~t,z~$ plane.
Namely, with a coordinate transformation $~\rho=\exp{(r)}~$, the solution
can be  rewritten as

$$
ds^{2}~=~-\sqrt{6}{e^{-\eta_{0}} \over Q} { dt^2 - dz^2 \over \ln \rho}
+ {2 d\rho^2 \over \Lambda (\rho^{2} - 1)^{2}}
+ e^{2\eta_{0}}\rho^{2}~d\phi^{2}
\eqno(32)
$$

\noindent and here the horizon is located at $~\rho=1~$. Still, due to the
presence of the logarythm in the metric, the solution is singular at both the
origin and infinity. The singularity at infinity should not come as a
surprise since, as argued in the previous section, the metric above has the
(quasi) Newtonian limit and the singularity at $~\infty~$ can be understood
as the standard infrared divergence.

For all other solutions, the ``spatial infinities'' are singular, as can be
seen from investigating the curvature invariants.
Hence, they all contain naked singularities.

Therefore, it has been demonstrated that the WZWN sigma model
construction of Raiten is the \underbar{unique}
axionically charged  cyllindrically symmetric target space of the $~4D~$
supergravity multiplet  in  string theory with central charge
deficit which does
not have naked singularities, i.e. obeys cosmic censorship!
The other solution which was found to be free of naked singularities
represents an entire space-time without singularities, but with
horizons. It corresponds to an
extremal black string solution.
It can be extended by gluing an identical copy resulting in a total space-time
where observer can travel from one region into the other.

In a sense, this result could be
dubbed the ``no-hair'' theorem for black strings, saying that the strings are
described by their mass, axion charge and  external
dilaton hair, and develop well
behaved horizons. We will not dwell here on the global properties of this
solution as it has  already been studied$^{15,16)}$. We remark however
that in order to obtain the full picture about the global properties oh the
solution, one should study geodesics of conformal transforms of the world
sheet metric. They describe motion of particles with different conformal
charges (scaling dimensions) and thus couple differently to
the dilaton$^{23)}$. The solution (28) is thence standing out as
motion of probes in it is universal, irrespective of the nature of test
particles, due to the constancy of the dilaton.

The string theory targets with no central charge deficit
are analyzed in exactly the same way.
The task is much easier here, though. Repeating the study of the
singularity structure of the solution (22) along the lines
outlined above, one can easily verify that none of these solutions contain a
point equivalent to $~r_{\infty}~$, and contain a singularity at the equivalent
of $~r_{s}~$. Furthermore, the ``spatial infinities'' there are also singular.
Thus, every such solution involves naked singularities.
So there are no black
strings among the string  theory target spaces that bear no central charge.

\vskip1.5cm
{\bf Summary}
\vskip1.5cm

Studies of the nonperturbative aspects of string theory have recently produced
novel, unorthodox solutions to the effective theory which appear to bear
resemblance  to the conventional black holes. They involve nontrivial
causal structure of the space-time, and feature horizons. Here general
solutions of the 4D effective theory for the supergravity multiplet
with cylindrically symmetric target spaces
have been studied. It was found that the only solutions consistent with
the (hoped for) cosmic censorship are indeed the gauged WZWN sigma models.
These solutions are just the 3D black hole of Horne and Horowitz, crossed
with a circle, which is necessary to carry the axion charge. In 4D it is given
by an integral of a one-form, and the subspace over which the integration is
performed must be compact to give rise to a finite charge.
They appear to be the unique space-time configurations which give rise to
event horizons and thus in a sense justify the trust of representing
universal theories that may be used to describe the string theory ground
states.

These solutions may also have relevance in cosmology,
as it is well known that there is a simple relationship between static
axially symmetric geometries and Bianchi I anisotropic universes. In that
case, the axion charge could be interpreted as a homogeneous distribution
of axion condensate in the universe, irregardles of whether the universe
had a compactified direction or not. Some of similar solutions were
considered recently by Tseytlin$^{24)}$.

\vskip1.5cm
{\bf Acknowledgements}
\vskip1.5cm

The author would like to thank K. Olive for helpful conversations.
This work has been supported in part by the University of Minnesota Doctoral
Dissertation Fellowship.
\vfil
\eject

{\bf References}
\vskip1.5cm
\item{[1]} A. Vilenkin, Phys. Rev. {\bf D23} (1981) 852; Phys. Rep. {\bf 121}
(1985) 263.
\item{[2]} D. Garfinkle, Phys. Rev. {\bf D32} (1985) 1323.
\item{[3]} E. Witten, {\bf Relativistic Astrophysics Proceedings, Chicago 1986}
(1986) 606.
\item{[4]} D. Harari and P. Sikivie, Phys. Rev. {\bf D37} (1988) 3438.
\item{[5]} A.G. Cohen and D.B. Kaplan, Phys. Lett. {\bf B215} (1988) 67.
\item{[6]} R. Gregory, Phys. Lett. {\bf B215} (1988) 663.
\item{[7]} D. Harari and A.P. Polychronakos, Phys. Lett. {\bf B240} (1990) 55.
\item{[8]} G.W. Gibbons, M.E. Ortiz and F. Ruiz Ruiz, Phys. Lett. {\bf B240}
(1990) 50.) 50.
\item{[9]} E. Witten, Phys. Lett. {\bf B153} (1982)
\item{[10]} B.R. Greene, A. Shapere, C. Vafa and S.-T. Yau, Nucl. Phys.
{\bf B337} (1990) 1.
\item{[11]} S. Ceccoti, Phys. Lett. {\bf B244} (1990) 23.
\item{[12]} C. Gundlach and M.E. Ortiz, Phys. Rev. {\bf D42} (1990) 2521.
\item{[13]} E. Witten, Phys. Rev. {\bf D44} (1991) 314.
\item{[14]} G.T. Horowitz and A. Strominger, Nucl. Phys. {\bf B350} (1991) 197.
\item{[15]} J.H. Horne and G.T. Horowitz, Nucl. Phys. {\bf B368} (1992) 444.
\item{[16]} E. Raiten, Fermilab preprint FERMILAB-PUB-91-338-T, Dec. 1991.
\item{[17]} P. Horava, Phys. Lett. {\bf B278} (1992) 101;
D. Gershon, Tel Aviv University preprint TAUP-1937-91, Dec 1991.
\item{[18]} S. K. Kar, S. P. Khastgir and G. Sengupta, IP Bhubaneswar
preprint IP/BBSR/92-35, May 1992; S. Mahapatra, Tata Institute preprint
92-28, May, 1992.
\item{[19]} I. Bars and K. Sfetsos, USC preprint USC-92/HEP-B1, May 1992;
I. Bars and K. Sfetsos, USC preprint USC-92/HEP-B2, May 1992;
K. Sfetsos, USC preprint USC-92/HEP-S1, June 1992;
\item{[20]} A. Dabholkar and J. Harvey, Phys. Rev. Lett {\bf 63} (1989) 719;
A. Dabholkar, G. Gibbonws, J. Harvey and F. Ruiz, Nucl. Phys.
{\bf 340} (1990) 33; Soo Jong Rey, Phys. Rev. {\bf D43} (1991) 526.
\item{[21]} B.A. Campbell, M.J. Duncan, N. Kaloper and K.A. Olive, Nucl. Phys.
{\bf B351} (1991) 778.
\item{[22]} J. Antoniades, C. Bachas, J. Ellis and D. Nanopoulos,
Phys. Lett. {\bf B221} (1988) {393; J. Antoniades, C. Bachas and A. Sagnotti,
Phys. Let. {\bf B235} (1990) 255.
\item{[23]} A. Shapere, S. Trivedi and F. Wilczek, IAS preprint
IASSNS-HEP-91/93, June 1991.
\item{[24]} A. A. Tseytlin, Cambridge University preprint, DAMTP-92-06,
June 92

\bye